\documentclass[10pt]{article} 
\usepackage[preprint]{tmlr}

\usepackage{amsmath,amsfonts,bm}

\def\eqref#1{equation~\ref{#1}}

\def\1{\bm{1}}

\DeclareMathAlphabet{\mathsfit}{\encodingdefault}{\sfdefault}{m}{sl}
\SetMathAlphabet{\mathsfit}{bold}{\encodingdefault}{\sfdefault}{bx}{n}

\usepackage{hyperref}
\usepackage{url}

\usepackage{tikz}
\usetikzlibrary{trees, positioning, shapes.geometric, arrows.meta, calc}
\usepackage{pgfplots}

\title{Content filtering methods for music recommendation: \\ A review}

\author{\name Terence Zeng \email terence6@illinois.edu \\
      \addr Department of Electrical and Computer Engineering\\
      University of Illinois Urbana-Champaign\\
      \AND 
      \name Abhishek K. Umrawal \email aumrawal@illinois.edu \\
      \addr Department of Electrical and Computer Engineering\\
      University of Illinois Urbana-Champaign
      }

\def\month{MM}  
\def\year{YYYY} 
\def\openreview{\url{https://openreview.net/forum?id=XXXX}} 

\begin{document}

\maketitle

\begin{abstract}
Recommendation systems have become essential in modern music streaming platforms, shaping how users discover and engage with songs. One common approach in recommendation systems is collaborative filtering, which suggests content based on the preferences of users with similar listening patterns to the target user. However, this method is less effective on media where interactions are sparse. Music is one such medium, since the average user of a music streaming service will never listen to the vast majority of tracks. Due to this sparsity, there are several challenges that have to be addressed with other methods. This review examines the current state of research in addressing these challenges, with an emphasis on the role of content filtering in mitigating biases inherent in collaborative filtering approaches. We explore various methods of song classification for content filtering, including lyrical analysis using Large Language Models (LLMs) and audio signal processing techniques. Additionally, we discuss the potential conflicts between these different analysis methods and propose avenues for resolving such discrepancies.
\end{abstract}

\section{Introduction}

Recommendations have played a central role in helping people discover new content for centuries. Traditionally, recommendations were made through word-of-mouth - a friend might suggest a good book they recently read, or recommend a new restaurant they dined at. Nowadays, however, the sheer abundance of available content in various media has made it impractical for users to rely solely on personal recommendations. One such medium is music. Spotify, the largest streaming service, has over 100 million songs. Users will struggle to manually discover new songs that are aligned with their preferences. To address this, music streaming services employ recommender systems, which suggest songs that users are likely to enjoy.

Before recommender systems became widely adopted, information retrieval served as a way to filter out vast amounts of content. Spearheaded largely by the foundational work of \cite{Rijsbergen}, these systems became widely used in areas such as library science, where the challenging of indexing and accessing large amounts of content required more efficient methods.

Recommender systems evolved from information retrieval systems as the demand for personalized access to information grew. While information retrieval systems rely on user queries to search for content, recommender systems offer relevant content by utilizing user behavior, eliminating the need for these queries.

The earliest recommender systems relied primarily on collaborative filtering \citep{Goldberg}, a technique that generates recommendations based on user-interaction data. Collaborative filtering works by identifying users with similar listening patterns and recommending items that those similar users enjoyed. This method saw widespread adoption by the early 2000s on platforms such as Amazon \citep{Linden} due to its simplicity and effectiveness. However, collaborative filtering has several limitations, particularly in the domain of music.

One such limitation is that user-interaction data is sparse. For large song databases, the sparsity can surpass 99.9\% \citep{Dror}, meaning that most users will have listened to less than 0.1\% of all songs. It becomes difficult for recommendation systems to accurately model user preferences when there is so little data about what they have or have not listened to. This sparsity not only limits the system's ability to find personalized recommendations but also reinforces popularity bias, as the system may disproportionately recommend widely-streamed tracks simply because they have more interaction data. This makes it especially difficult for lesser-known artists to gain visibility, ultimately reducing the diversity of recommendations.

This problem is exacerbated by the fact that explicit feedback is rarely provided in music. In other media, such as movies, a numerical rating is assigned by a user to an item. Music streaming services generally lack this functionality - the main types of feedback are a user's listening history and playlists. In fact, even this implicit data isn't reliable - there are many reasons why someone may listen to a song they don't completely enjoy. Therefore, we decided to explore content-based filtering, which examines audio characteristics rather than user-item interactions. 

One of the earliest frameworks for audio analysis was MARSYAS, developed by \cite{Tzanetakis}. Inspired by the aforementioned information retrieval techniques, the MARSYAS framework audio extracts some information from the raw signals and stores it in a structured text format, allowing for traditional retrieval techniques to be used. A key innovation of MARSYAS was its approach to audio representation: rather than storing audio as raw waveform data, it transformed it into the frequency domain, giving access to powerful features such as spectral centroid and Mel-frequency Cepstral Coefficients (MFCCs). Some of these features were proven to be effective by the work of \cite{davis}. These features were then used in supervised and unsupervised learning models to perform tasks like emotion recognition, genre classification, and instrument detection. By bridging ideas from information retrieval and audio signal processing, MARSYAS laid the foundation for many future content-based music recommendation systems.

\cite{Logan} further explored MFCCs, using them to filter music based on content. Their work was motivated by the growing need to create playlists with similar-sounding songs, since digital media consumption was moving from albums to individual tracks. To find similar-sounding songs, they used frequency analysis combined with k-means clustering within each song. These clusters represent the musical shape of each song. With this system, the proportion of similar-sounding songs was much higher than with a randomly generated playlist. These methods and results showed that audio analysis was an area worth looking at and paved the way for more sophisticated methods.

\begin{figure}[h!]
\centering
\begin{tikzpicture}[
    box/.style={rectangle, draw=gray!60, fill=gray!5, thick, minimum width=3.5cm, minimum height=1cm, align=center, rounded corners},
    every node/.style={font=\small},
    node distance=2.5cm
]

\node[box] (ir) {Information Retrieval};
\node[box, right=1cm of ir] (audio) {Audio Analysis};
\node[box, right=1cm of audio] (music) {Music Similarity};

\node[above=0.2cm of ir] {\cite{Rijsbergen}};
\node[above=0.2cm of audio] {\cite{Tzanetakis}};
\node[above=0.2cm of music] {\cite{Logan}};

\draw[thick, yellow!60, fill=yellow!5] (5,-3.5) ellipse (5cm and 2cm);
\draw[thick, orange!60, fill=orange!5] (5,-4.25) ellipse (2.5cm and 1.25cm);
\draw[thick, cyan!60, fill=cyan!5] (5,-4.75) ellipse (1.5cm and 0.75cm);

\node[text width=4cm, align=center] at (5, -2.25) {Multimedia \\ (video, audio, text)};
\node at (5, -3.5) {Audio};
\node at (5, -4.75) {Music};
\node at (-3, 0) {\textbf{Methods}};
\node at (-1.5, -3.5) {\textbf{Domains}};

\draw[->, thick] (ir) -- (audio);
\draw[->, thick] (audio) -- (music);

\draw[->, thick] (ir) -- (1, -2.3);
\draw[->, thick] (3, -0.5) -- (3, -3.5);
\draw[->, thick] (music) -- (6.3, -4.4);

\end{tikzpicture}
\caption{Timeline of the development of music similarity analysis.}
\end{figure}
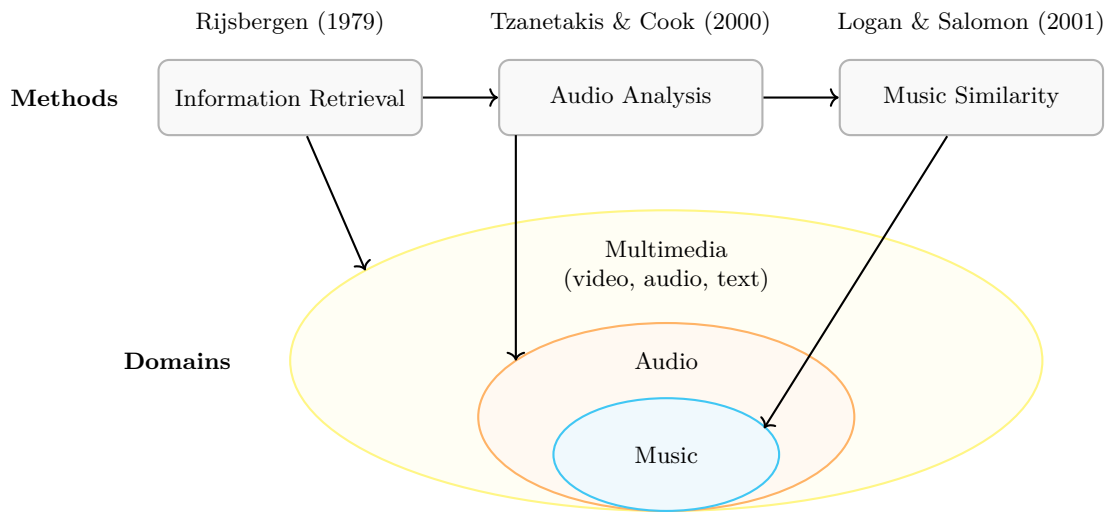

In recent years, as music streaming platforms have scaled to host tens of millions of songs, the need for effective and personalized recommendation systems has become increasingly central. This has driven a shift from early collaborative filtering approaches to more sophisticated, content-aware techniques that leverage audio signals, lyrics, emotional context, and user demographics. These advances are largely fueled by progress in deep learning, signal processing, and natural language processing, especially with the introduction of large language models (LLMs) and hybrid multi-modal systems.

\vspace{50 pt}

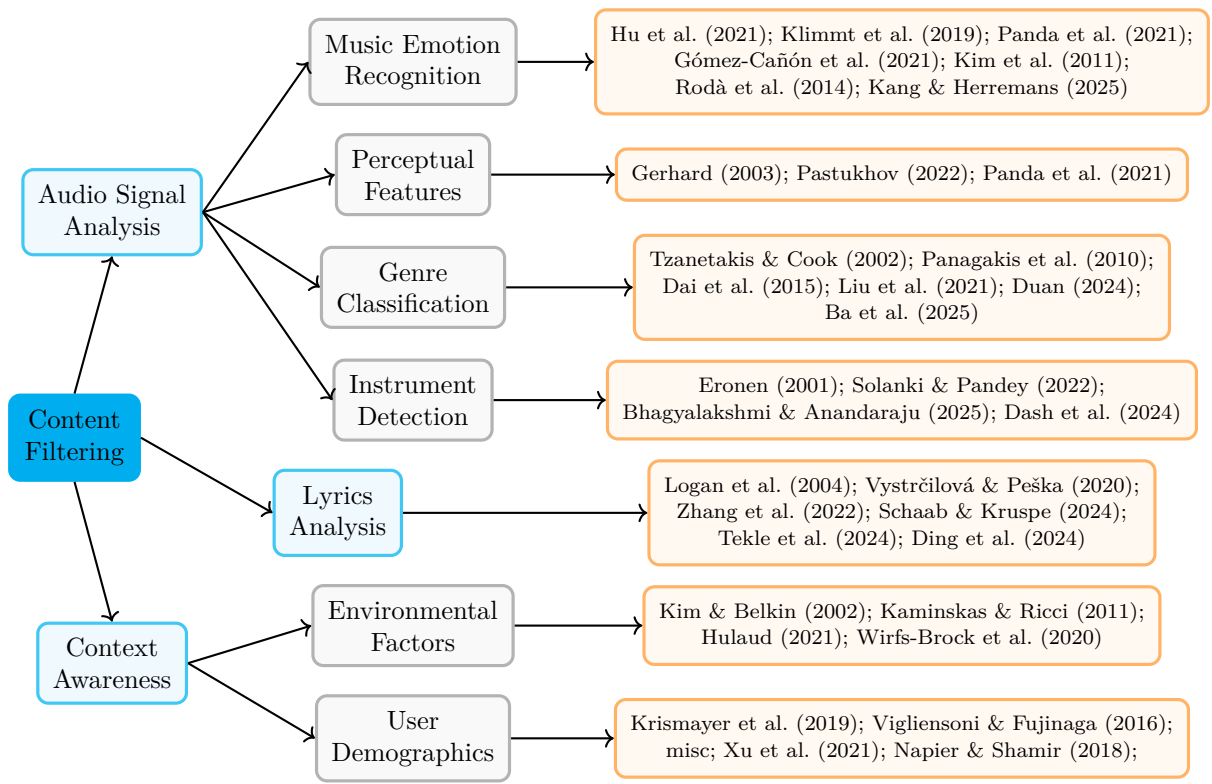
\begin{figure}[h!]
    \centering
\begin{tikzpicture}[
Title/.style={rectangle, draw=cyan!100, fill=cyan!100, very thick, inner sep=2mm, align=center, rounded corners},
Heading/.style={rectangle, draw=cyan!60, fill=cyan!5, very thick, inner sep=2mm, align=center, rounded corners},
Normal/.style={rectangle, draw=gray!60, fill=gray!5, very thick, inner sep=2mm, align=center, rounded corners},
Citation/.style={rectangle, draw=orange!60, fill=orange!5, very thick, inner sep=2mm, align=center, rounded corners, font=\fontsize{8}{10}\selectfont}
]
\node[Title] (ContentFiltering) at (0.5, 0) {Content \\ Filtering};
\node[Heading] (Audio) at (1, 3) {Audio Signal \\ Analysis};
\node[Heading] (Lyrics) at (4, -1) {Lyrics \\ Analysis};
\node[Heading] (Context) at (1, -3) {Context \\ Awareness};

\node[Normal] (MER) at (5, 5) {Music Emotion \\ Recognition};
\node[Normal] (Perceptual) at (5, 3.5) {Perceptual \\ Features};
\node[Normal] (Genre) at (5, 2) {Genre \\ Classification};
\node[Normal] (Instrument) at (5, 0.5) {Instrument \\ Detection};
\node[Normal] (Environmental) at (5, -2.5) {Environmental \\ Factors};
\node[Normal] (Demographics) at (5, -4) {User \\ Demographics};

\node[Citation] (MER_Cite) at (11.5, 5) {\cite{HuXiao}; \cite{Klimmt}; \cite{Panda}; \\ \cite{Gomez}; \cite{Kim}; \\ \cite{Roda}; \cite{Kang}};
\node[Citation] (Perceptual_Cite) at (11.5, 3.5) {\cite{gerhard}; \cite{pastukhov}; \cite{Panda}};
\node[Citation] (Genre_Cite) at (11.5, 2) {\cite{tzanetakis_genre}; \cite{panagakis}; \\ \cite{dai}; \cite{liu}; \cite{duan}; \\ \cite{ba}};
\node[Citation] (Instrument_Cite) at (11.5, 0.5) {\cite{eronen}; \cite{solanki}; \\ \cite{bhagyalakshmi}; \cite{dash}};
\node[Citation] (Lyrics_Cite) at (11.5, -1) {\cite{logan_semantic}; \cite{vystrvcilova}; \\ \cite{zhang}; \cite{schaab}; \\ \cite{tekle}; \cite{ding}};
\node[Citation] (Environmental_Cite) at (11.5, -2.5)
{\cite{kim_context}; \cite{kaminskas}; \\ \cite{Hulaud}; \cite{wirfs}};

\node[Citation] (Demographics_Cite) at (11.5, -4) { \cite{krismayer}; \cite{vigliensoni}; \\
misc; \cite{xu}; \cite{napier};
};

\draw[->, thick] (ContentFiltering.north) to (Audio.south);
\draw[->, thick] (ContentFiltering.east) to (Lyrics.west);
\draw[->, thick] (ContentFiltering.south) to (Context.north);

\draw[->, thick] (Audio.east) to (MER.west);
\draw[->, thick] (Audio.east) to (Perceptual.west);
\draw[->, thick] (Audio.east) to (Genre.west);
\draw[->, thick] (Audio.east) to (Instrument.west);

\draw[->, thick] (Context.east) to (Environmental.west);
\draw[->, thick] (Context.east) to (Demographics.west);

\draw[->, thick] (Lyrics.east) to (Lyrics_Cite.west);

\draw[->, thick] (MER.east) to (MER_Cite.west);
\draw[->, thick] (Perceptual.east) to (Perceptual_Cite.west);
\draw[->, thick] (Genre.east) to (Genre_Cite.west);
\draw[->, thick] (Instrument.east) to (Instrument_Cite.west);
\draw[->, thick] (Environmental.east) to (Environmental_Cite.west);
\draw[->, thick] (Demographics.east) to (Demographics_Cite.west);

\end{tikzpicture}
\caption{An overview of the most common methods used in content filtering and the respective papers.}
\end{figure}

\vspace{50 pt}

\section{Audio Signal Analysis}

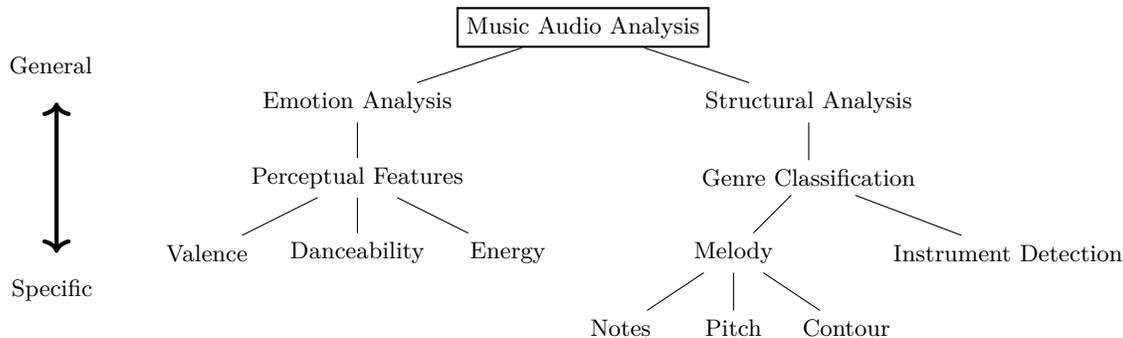
\begin{figure}[h!]
\centering
\begin{tikzpicture}[
  level 1/.style={sibling distance=60mm},
  level 2/.style={sibling distance=20mm},
  level 3/.style={sibling distance=20mm},
  level 4/.style={sibling distance=15mm},
  level distance=10mm,
  every node/.style = {align=center, font=\small},
  edge from parent/.style={draw}
  ]

\coordinate (origin) at (-7,-1);
\draw[<-> , ultra thick] (origin) -- ++(0,-2);
\node[anchor=east] at (-6.4,-0.5) {General};
\node[anchor=east] at (-6.4,-3.5) {Specific};

\node[draw, thick] {Music Audio Analysis}
  child { node {Emotion Analysis}
    child { node {Perceptual Features}
      child { node {Valence}}
      child { node {Danceability}}
      child { node {Energy}}
    }
  }
  child { node {Structural Analysis}
    child { node {Genre Classification}
      child { node {Melody}
        child { node {Notes} }
        child { node {Pitch} }
        child { node {Contour} }
      }
      child { node [anchor=west] {Instrument Detection}}
    } 
  };

\end{tikzpicture}
\caption{A hierarchical taxonomy of music analysis.}
\end{figure}

\subsection{Music Emotion Recognition}

Music Emotion Recognition (MER) plays a crucial role in modern music recommendation systems by enabling more personalized and context-aware suggestions. By incorporating MER, recommendation systems can adapt to a user’s mood, activities, or listening context, offering more meaningful and emotionally resonant suggestions. This adaptation is particularly beneficial in applications like mood-based playlist generation. Features such as tempo, mode, timbre, and harmonic structure are commonly used to infer emotional content, while natural language processing (NLP) techniques analyze lyrics for sentiment and thematic elements. 

Experiments have shown that music can convey certain emotions in people and alter their behaviors. In a study of university students, 85\% of participants perceived that listening to music has positive emotional effects \citep{HuXiao}.  Additionally, in a study of 68 video game players by \cite{Klimmt}, a significant increase (p < 0.01) in player enjoyment was observed once music was included. We also know that music categorization is essential to streaming platforms, and Spotify, for example, has spent significant money and research effort on this area \citep{Panda}. Thus, there is significant motivation to improve the emotional classification of music. 

There are two predominant types of music taxonomies - ways in which music is emotionally classified \citep{Gomez}. The first type is categorical, where music is separated into distinct classes (Figure \ref{Categorical}). Just as music is often separated into genres (e.g., pop, metal, rap), music can also be separated into emotional effects (e.g., happy, angry, excited). While this approach is natural and easy to understand, it doesn't capture the full range of emotions in music. On the other hand, a dimensional approach measures music on multiple scales. Most commonly, mood is measured by the two dimensions of arousal and valence, first proposed by \cite{russell}. \cite{Kim} developed the idea of using this framework to classify music (Figure \ref{Dimensional}). Different dimensions are also being explored, such as potency, tension, and energy \cite{Roda}. This dimensional approach allows for a wider and more diverse range of emotions. However, it is not intuitive for users to qualify songs based on these dimensions. While mainstream platforms primarily use categorical classification because of its communicability, hybrid approaches are being researched and will likely surpass a categorical approach in effectiveness \citep{Kang}.

\vspace {12pt}

\begin{figure}[htbp]
    \centering
    \begin{minipage}{0.5\textwidth}
        \begin{tikzpicture}
            \begin{axis}[axis lines=left, xlabel=valence, ylabel=arousal, xticklabels=\empty, yticklabels=\empty, 
            xmin = 0, xmax = 0.6, ymin = 0, ymax = 0.6,
            xtick = {0, 0.2, 0.4, 0.6},
            grid=both]
            \node at (axis cs:0.1, 0.1) {Sad};
            \node at (axis cs:0.1, 0.3) {Distressed};
            \node at (axis cs:0.1, 0.5) {Angry};
            \node at (axis cs:0.3, 0.5) {Stimulating};
            \node at (axis cs:0.5,0.5) {Excited};
            \node at (axis cs:0.5, 0.3) {Happy};
            \node at (axis cs:0.5, 0.1) {Relaxed};
            \node at (axis cs:0.3, 0.1) {Tired};
            \addplot[only marks, mark=*, color=red] coordinates {(0.38, 0.41)};
            \end{axis}
        \end{tikzpicture}
        \caption{In the dimensional representation, emotions are defined on two continuous axes. Each labeled region represents a general emotional category based on position in the space. The red dot marks an arbitrary song, located based on its valence and arousal scores.} \label{Dimensional}
    \end{minipage}%
    \hspace{20pt} 
    \begin{minipage}{0.45\textwidth}
        \begin{tikzpicture}[
        Center/.style={circle, draw=red!100, fill=red!100, scale = 0.45},
        Heading/.style={rectangle, draw=gray!60, inner sep=2mm, align=center},
        Target/.style={rectangle, draw=cyan!60, inner sep=2mm, align=center}
        ]
        \node[Center] (center) at (0, 0){};
        \node[Heading] (Excited) at (2.5, 2) {Excited};
        \node[Heading] (Happy) at (2.5, 0) {Happy};
        \node[Heading] (Relaxed) at (2.5, -2) {Relaxed};
        \node[Heading] (Tired) at (0, -3) {Tired};
        \node[Heading] (Sad) at (-2.5, -2) {Sad};
        \node[Heading] (Distressed) at (-2.5, -0) {Distressed};
        \node[Heading] (Angry) at (-2.5, 2) {Angry};
        \node[Target] (Stimulating) at (0, 3) {Stimulating};

        \draw[->, thick] (center.north east) to (Excited.south west);
        \draw[->, thick] (center.east) to (Happy.west);
        \draw[->, ultra thin] (center.south east) to (Relaxed.north west);
        \draw[->, ultra thin] (center.south) to (Tired.north);
        \draw[->, ultra thin] (center.south west) to (Sad.north east);
        \draw[->, ultra thin] (center.west) to (Distressed.east);
        \draw[->, ultra thin] (center.north west) to (Angry.south east);
        \draw[->, ultra thick] (center.north) to (Stimulating.south);
        \end{tikzpicture}
        \caption{In the categorical representation, emotions are modeled as categories, surrounding the same song as in Figure \ref{Dimensional}. Each arrow points to a possible emotional label, with arrow widths representing the suitability of the label. Based off the song's features, the label \textit{Stimulating} is selected as the most appropriate classification.} \label{Categorical}
    \end{minipage}
\end{figure}

As music consumption becomes increasingly personalized, MER will continue to evolve, integrating deep learning models and multimodal data to enhance the emotional intelligence of recommendation systems \citep{Hao}. Recently, LLMs have employed MER to generate music based on user prompts \citep{Yuan}. In the near future, we can expect these models to become more sophisticated.

\subsection{Perceptual Features}

A large part of music emotion recognition is fundamentally based on perceptual features. These are quantifiable attributes that serve as a middle ground between broad features such as genre of the piece and fine details such as pitch of a note \citep{gerhard}. 

Spotify uses twelve perceptual features as part of its music recommendation pipeline. Danceability, energy, and valence are some of these perceptual features \citep{pastukhov}. Danceability, a perceptual feature that is intuitively understandable, draws from the more cryptic but objective details such as rhythm stability and beat strength. Energy, another term that is intuitively understandable, observes dynamic range, loudness, and general entropy.

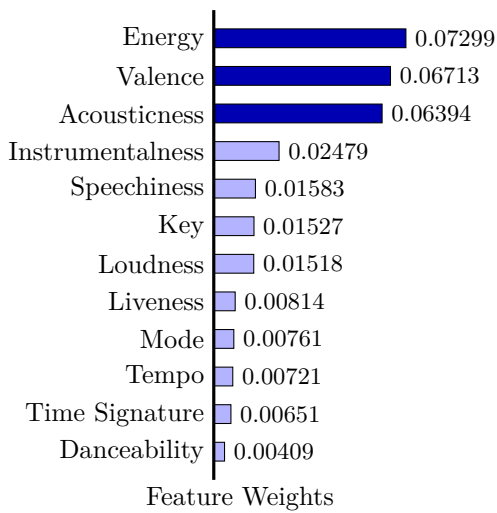
\begin{figure}[h]
\centering
\begin{tikzpicture}[xscale=35, yscale=0.5] 

\draw[fill=blue!70!black] (0,11) rectangle (0.07299,11.5);
\draw[fill=blue!70!black] (0,10) rectangle (0.06713,10.5);
\draw[fill=blue!70!black] (0,9) rectangle (0.06394,9.5);

\draw[fill=blue!30] (0,8) rectangle (0.02479,8.5);
\draw[fill=blue!30] (0,7) rectangle (0.01583,7.5);
\draw[fill=blue!30] (0,6) rectangle (0.01527,6.5);
\draw[fill=blue!30] (0,5) rectangle (0.01518,5.5);
\draw[fill=blue!30] (0,4) rectangle (0.00814,4.5);
\draw[fill=blue!30] (0,3) rectangle (0.00761,3.5);
\draw[fill=blue!30] (0,2) rectangle (0.00721,2.5);
\draw[fill=blue!30] (0,1) rectangle (0.00651,1.5);
\draw[fill=blue!30] (0,0) rectangle (0.00409,0.5);

\node[anchor=east] at (0,11.25) {Energy};
\node[anchor=east] at (0,10.25) {Valence};
\node[anchor=east] at (0,9.25) {Acousticness};
\node[anchor=east] at (0,8.25) {Instrumentalness};
\node[anchor=east] at (0,7.25) {Speechiness};
\node[anchor=east] at (0,6.25) {Key};
\node[anchor=east] at (0,5.25) {Loudness};
\node[anchor=east] at (0,4.25) {Liveness};
\node[anchor=east] at (0,3.25) {Mode};
\node[anchor=east] at (0,2.25) {Tempo};
\node[anchor=east] at (0,1.25) {Time Signature};
\node[anchor=east] at (0,0.25) {Danceability};

\node[anchor=west, font=\small] at (0.07299,11.25) {0.07299};
\node[anchor=west, font=\small] at (0.06713,10.25) {0.06713};
\node[anchor=west, font=\small] at (0.06394,9.25) {0.06394};
\node[anchor=west, font=\small] at (0.02479,8.25) {0.02479};
\node[anchor=west, font=\small] at (0.01583,7.25) {0.01583};
\node[anchor=west, font=\small] at (0.01527,6.25) {0.01527};
\node[anchor=west, font=\small] at (0.01518,5.25) {0.01518};
\node[anchor=west, font=\small] at (0.00814,4.25) {0.00814};
\node[anchor=west, font=\small] at (0.00761,3.25) {0.00761};
\node[anchor=west, font=\small] at (0.00721,2.25) {0.00721};
\node[anchor=west, font=\small] at (0.00651,1.25) {0.00651};
\node[anchor=west, font=\small] at (0.00409,0.25) {0.00409};

\draw[very thick] (0,-0.5) -- (0,12);

\node at (0.01, -1.0) {Feature Weights};

\end{tikzpicture}
\caption{Usefulness of perceptual features to discriminate songs on arousal-valence axes, adapted from \cite{Panda}}
\label{spotify_perceptual}.
\end{figure}

The effectiveness of these perceptual features appears to be highly context-dependent, but can potentially be useful for music emotion recognition \citep{Panda}. In particular, \textit{energy}, \textit{valence}, and \textit{acousticness} seem to effectively discriminate songs across the valence and arousal axes, as shown in Figure \ref{spotify_perceptual}. Naturally, \textit{energy} and \textit{valence} are the definitions of the axes themselves. More notably, however, \textit{acousticness} has a strong negative correlation with arousal, suggesting that songs with more electronic tones are higher in energy. 

By leveraging these features, music streaming platforms can combine attributes that are very objective - tempo, timbre, etc. - into features that are relevant for music emotion recognition. This layered approach allows systems to bridge the gap between low-level signal analysis and human emotional perception, enabling more personalized and emotionally relevant recommendations. Furthermore, perceptual features can serve as a crucial foundation for hybrid recommendation systems that integrate content-based and context-aware approaches, providing a richer understanding of how music impacts listeners across different moods, activities, and environments.

\subsection{Genre Classification}

While perceptual features describe music at a relatively granular level, they can also be aggregated to classify music into broader categories. One of the most common ways this is done is through genre classification, a task that has long played a central role in organizing and recommending music.

Genre classification is a foundational task in music information retrieval and recommendation. Genres serve as high-level labels that group music based on stylistic and cultural similarities, such as pop, jazz, classical, or hip-hop. Genres are both easily understandable and communicable to the average user in a way that perceptual features may not be. From a signal processing standpoint, genre classification often involves extracting features that reflect rhythmic patterns, spectral content, and timbral characteristics of a track.

Early methods relied on manually engineered features and classifiers such as k-nearest neighbors and Gaussian mixture models (GMMs). \cite{tzanetakis_genre} applied these classifiers to feature sets that were related to the perceptual features of timbre, rhythm, and pitch. The GTZAN dataset was also developed by \cite{tzanetakis_genre} for the purpose of testing these methods, consisting of 1,000 labeled music samples spanning ten genres---this dataset is still commonly used today. Notably, the system was approximately as accurate as humans, achieving an accuracy of over 61\% for this dataset. This laid the foundation for more sophisticated classification schemes.

\cite{panagakis} made a significant leap in classification accuracy by utilizing audio temporal modulations. These methods aimed to mimic how humans process sound, significantly improving the accuracy to 91\%.

\cite{dai} utilized a special variation of deep neural networks (DNNs), raising the accuracy to 93.4\%. DNNs, which would normally be trained on one genre each, could be additionally trained on similar genres, defined by k-nearest neighbors. This gives each DNN more training data to work with. These methods were borrowed from the idea of ``multilingual'' DNNs, which used data from multiple languages to learn linguistic patterns.

\cite{liu} utilizes convolutional neural networks (CNNs), using visual representations of music, such as a spectrogram, to distinguish music genres. Previous implementations borrowed architectures from CNNs used in image recognition. Although these implementations reached acceptable accuracies, they were observed to learn features that were not necessarily the most suitable for music genre classification. \cite{liu} improved upon these naive implementations by taking into account time-frequency information, achieving a 93.9\% accuracy. \cite{duan} improved methods used to analyze spectrograms, focusing on low-level information to substantially increase the accuracy to 99.0\%. 

\cite{ba} employed capsule neural networks, an improvement on neural networks that grouped neurons. This allowed for a more holistic view on how finer patterns related to broader patterns, raising the accuracy to 99.9\%.

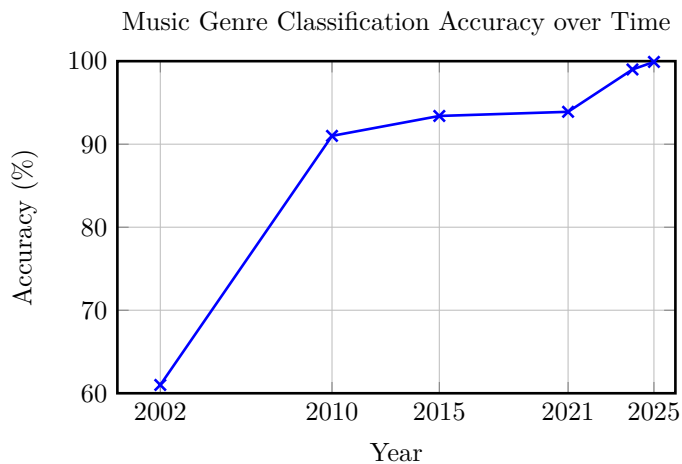
\begin{figure}[h]
\centering
\begin{tikzpicture}
\begin{axis}[
    /pgf/number format/1000 sep={},
    width=9cm,
    height=6cm,
    grid=both,
    title=Music Genre Classification Accuracy over Time,
    xlabel={Year},
    ylabel={Accuracy (\%)},
    ymin=60, ymax=100,
    xmin=2000, xmax=2026,
    xtick={2002, 2010, 2015, 2021, 2025},
    ytick={60, 70, 80, 90, 100},
    mark size=3pt,
    line width=1pt,
]
\addplot[
    color=blue,
    mark=x,
]
coordinates {
    (2002, 61)
    (2010, 91)
    (2015, 93.4)
    (2021, 93.9)
    (2024, 99.0)
    (2025, 99.91)
};
\end{axis}
\end{tikzpicture}
\caption{Progression of Music Genre Classification. This graph highlights milestones mentioned in the text above and does not include all improvements.}
\end{figure}

Genre classification not only helps users discover music in familiar categories but also serves as a proxy for emotional content, instrumentation, and even cultural context. In practice, it is often combined with other forms of content-based filtering to enhance personalization. In the future, genre classification may become more fluid and personalized to each individual, as we are able to get more user data.

\subsection{Instrument Detection}

While genres provide a broad stylistic overview, a more granular analysis involves understanding the specific instruments used in a track. Instrument detection adds another layer of interpretability to content-based filtering by revealing the texture and arrangement of a piece of music. The perceptual feature of acousticness, for instance, has a strong negative correlation with the arousal dimension. Therefore, detecting the instruments used is not only useful as a feature in itself, but also as a feature in MER systems.

Traditional approaches used spectral analysis and feature extraction to model instrument timbres. \cite{eronen} pioneered these methods, comparing different feature sets and their effectiveness in distinguishing instruments. Mel-frequency cepstral coefficients, which were developed earlier by \cite{rabiner} for speech recognition, were shown to be the most effective features. \cite{eronen} was able to achieve a 35\% accuracy for recognizing individual instruments and a 77\% accuracy for recognizing instrument families from a database of over 5000 tones played with a single instrument.

However, a significant challenge is that most music is polyphonic, consisting of multiple instruments played simultaneously and sharing overlapping frequency ranges. Multi-label classification is thus a desired goal of instrument recognition systems. \cite{solanki} employed CNNs to recognize multiple different instruments, achieving a 92.8\% accuracy on the annotated IRMAS (Instrument Recognition in Musical Audio Signals) dataset.

Knowing which instruments are present can aid in genre classification, emotion modeling, and even playlist generation. For example, acoustic guitars and pianos are commonly associated with calming or reflective music, while electric guitars and drums might align with high-energy genres like rock or metal.

Instrument detection not only enhances the interpretability of music for recommendation systems but also opens the door to advanced applications like automatic music transcription and audio source separation.

\section{Lyrics Analysis}

While audio features can capture some properties of a track, lyrics provide more direct insight into the semantic content of a song. Analysis of lyrics is valuable for MER and can improve music recommendations. Unlike audio signal processing, analyzing lyrics is computationally lightweight, offering a lower processing load while still capturing meaningful insights \cite{vystrvcilova}. This efficiency is especially important at scale, when recommender systems are in operation for thousands of users at the same time. 

With the advent of Large Language Models (LLMs), models can now analyze text at a much more nuanced level than before. These models can process long passages of text, understand context, and extract nuanced meaning---capabilities that were difficult or impossible with earlier NLP methods. Instead of relying on word frequency or fixed sentiment lexicons, LLMs can understand figurative language, recognize changing emotional tone across verses, and identify themes that extend beyond individual lines.

\cite{logan_semantic} conducted foundational research on lyrics analysis, using lyrics to determine genres of songs as well as artist similarities. Text was analyzed using probabilistic latent semantic analysis, finding meaning from words that were most unique to each track, such as ``love'' or ``time'', while ignoring more common words, such as ``the''. While genre classification had been developed in the past, this was the first notable attempt to use textual content to classify genres. Although the resulting accuracy was not as high as that of models that used purely audio analysis, this research showed that there are significant patterns to be found in song lyrics.

More recently, researchers have begun to explore hybrid models that combine audio and lyrical features. \cite{zhang} combined some aspects of audio analysis with lyrics analysis, pre-training models on audio samples to generate better interpretations of lyrics. This works because the emotional tone of the audio adds a layer of context on top of the text. Similarly, \cite{schaab} used both audio and lyrics to place songs on the arousal-valence plane.

\cite{tekle} experimented with using LLM-generated lyrics summaries as features in a recommendation system. Rather than inputting full lyrics, which can be lengthy or copyright restricted, this approach condenses the lyrical content of the song. These summaries offer a computationally efficient and less problematic way to integrate lyric-based emotion or theme analysis into tasks like genre classification or playlist generation. Their results support the idea that even short summaries, when generated by LLMs, retain enough semantic richness to be useful for music recommendation.

In the future, as we understand more about lyrics analysis, we may also be better positioned to generate new songs with meaningful lyrics. Extensive research has already been conducted in this area, such as the work of \cite{ding}, who showed that LLMs can generate lyrics conditioned on mood or genres. This opens up vast opportunities in AI-generated music, where a solid understanding of lyrics is essential.

\section{Context Awareness}

\subsection{Environmental Factors}

While many music recommendation systems draw from solely a user's listening history, more sophisticated systems incorporate the user's context as well. These situational components can include time of day, mood, and user activity. \cite{kim_context} showed that users often associate songs with certain emotions or events, rather than strictly with genres or artists. This suggests that incorporating contextual awareness into recommender systems can better align recommendations with the user’s emotional or situational needs. For example, music that is appropriate for a morning commute may differ significantly from that suited for a study session or a workout. This idea is especially relevant in shared or public spaces, such as restaurants, gyms, or retail stores, where the goal is often to evoke a particular atmosphere and ambience. Context-aware recommendation systems can dynamically adjust music to maintain a target mood or energy level across different times of day and social settings.

Early research helped lay the foundation for such systems. For example, \cite{kaminskas} developed a system that can select appropriate music for a context, similarly to how a real person would select such music. Their system calculated the similarity of songs to various places of interest according to feature-based descriptions. This work is notable because songs and places were previously thought of as difficult to match, as they seem unrelated.

\begin{figure}[h!]
    \centering
    \includegraphics[width=315pt]{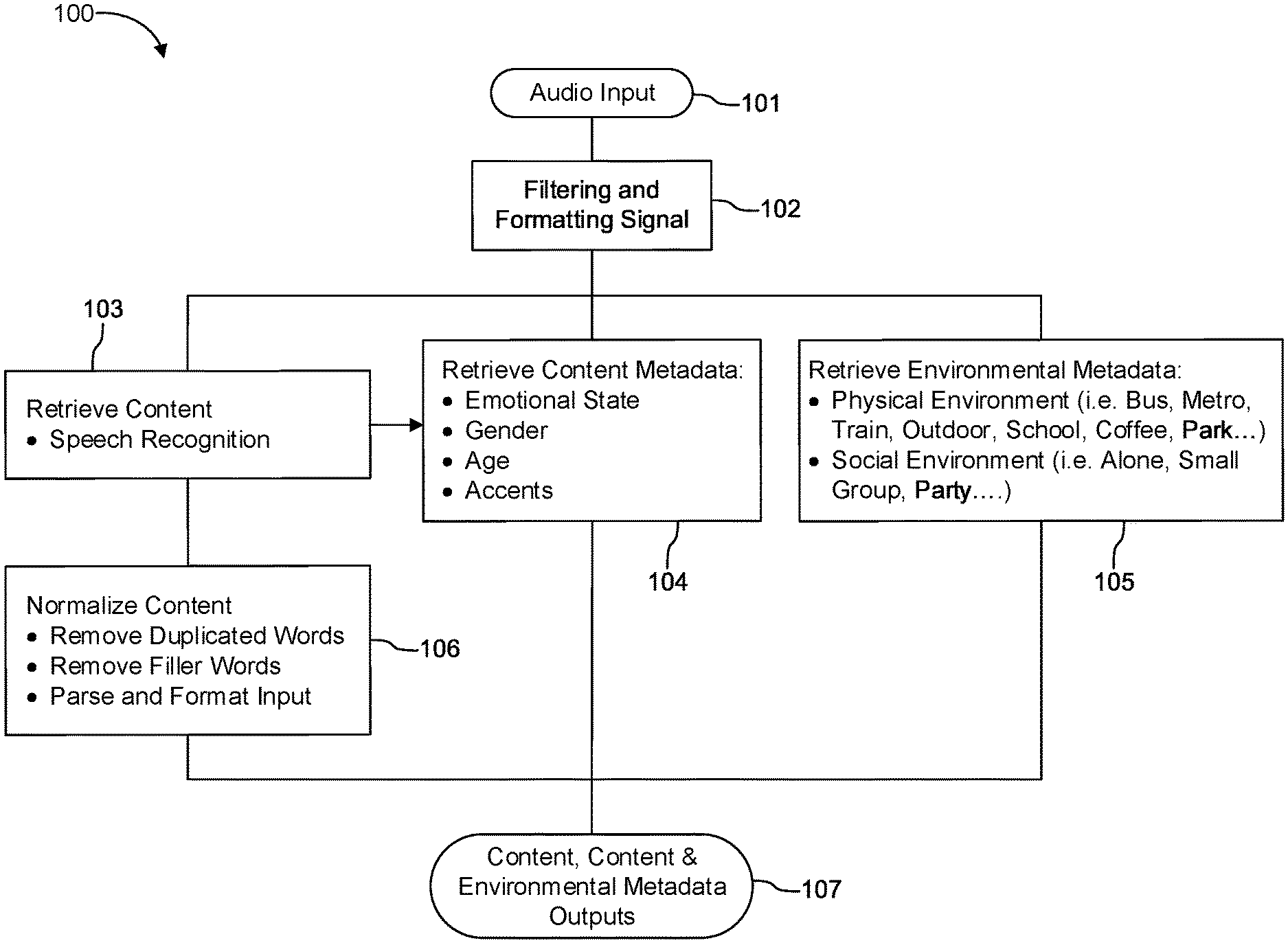}
    \caption{Spotify patent by \cite{Hulaud}; ``Identification of taste attributes from an audio signal''. Elements 104 and 105 indicate the retrieval of emotional state and environmental data through filtering and formatting the audio input.}
    \label{spotify_patent}
\end{figure}

In parallel, commercial platforms have also recognized the value of context. Spotify, for example, has filed patents showing its interest in using context for recommendations. One such patent by \cite{Hulaud} (Figure \ref{spotify_patent}) depicts an idea for detecting the user's environment and emotional state solely from audio input. For instance, by detecting background chatter or the cadence of the user's voice, the system might determine whether the user is in a car, at a party, or feeling anxious. These inferences could then shape real-time music recommendations, making the listening experience more responsive and adaptive. This type of system reflects a broader trend in recommendation research: designing systems that are not just personalized, but also sensitive to the user's current situation.

Researchers from industry and academia are collaborating to explore the ideas of using context-aware recommender systems to further personalize the music listening experience. For instance, \cite{wirfs} explored the idea of developing voice assistants into ``music coaches'', capable of utilizing user context to intelligently help listeners figure out what they want to hear. This concept envisions music recommendation as a dynamic, interactive process rather than a single-step process where some input generates some output. By leveraging voice input and contextual signals, a music coach could guide users through a process where users can articulate their current needs and emotions more effectively.

\subsection{User Demographics}

In addition to environmental context, user demographics---such as age, gender, and cultural background---play a pivotal role in shaping music preferences. These characteristics often influence not just the genres or artists a person listens to, but also their openness to musical exploration, emotional resonance with lyrics, and even the types of social experiences they associate with music. By incorporating demographic data into music recommendation systems, researchers and developers can design more personalized recommendation models.

Several studies have explored the relationship between listening history and user demographics. For example, \cite{krismayer} showed that age, gender, and nationality of a user can be predicted with reasonable accuracy from a user's listening profile. This finding suggests that there are consistent patterns in music consumption that align with demographic traits---patterns that can be leveraged for tailored music recommendations, even when explicit demographic data is unavailable.

Building upon this insight, some researchers have experimented with methods for integrating demographic data directly into recommendation algorithms. For instance, \cite{vigliensoni} showed that different age groups have significantly different listening behaviors. Younger users tend to stick to more mainstream users compared to older users. Additionally, users tend to be biased towards artists of the same gender as them, though this bias diminishes with age. Understanding these behaviors can help systems present users with options that are better aligned with their preferences.

\begin{figure}[h!]
    \centering
    \includegraphics[width=330pt]{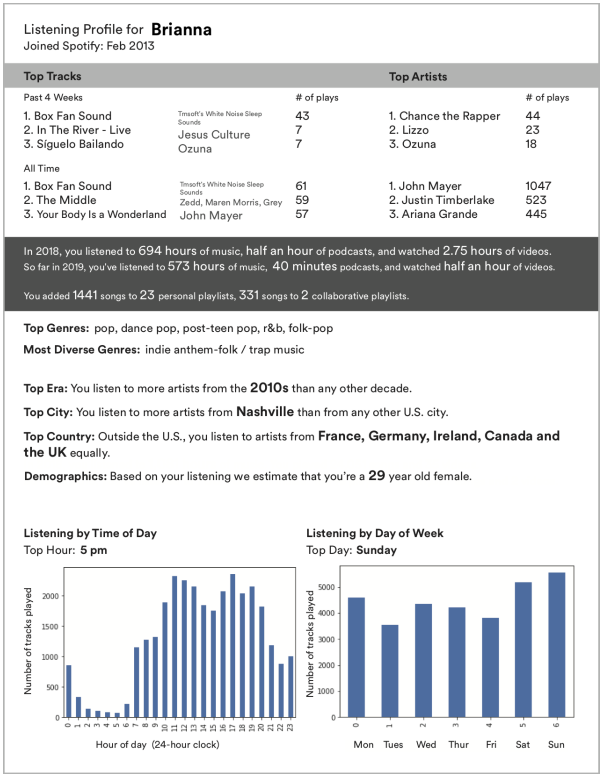}
    \caption{Visualization of how Spotify finds patterns in listening history and utilizes user demographics, from \cite{wirfs}.}
    \label{spotify_demographics}
\end{figure}

Commercial platforms have also begun to explore the use of inferred demographics. Spotify, for example, is researching systems for estimating demographic information based on a user's listening behavior (Figure \ref{spotify_demographics}). These estimates can be useful since commercial platforms often have limited or inaccurate data concerning users.

In the future, it is possible that sophisticated systems will detect user demographics from images in real time, such as in the work of \cite{mammadli}, which uses CNNs. In public settings such as gyms or cafes, this approach could dynamically adjust music to suit the general demographics of audiences.

Ultimately, incorporating demographic into recommendation systems not only enhances accuracy but also promotes inclusivity and fairness, which can improve exposure of newer, under-appreciated artists. Addressing the diverse backgrounds and cultural identities of users will be crucial in delivering meaningful experiences.

\section{Conclusion}

Content-based filtering has evolved into a rich and multifaceted field that goes far beyond matching audio similarity. This review has explored how music emotion recognition, perceptual features, lyrics analysis, context awareness, and instrument detection each contribute to creating more personalized and emotionally intelligent recommendation systems. These components interact in complex ways: perceptual features like valence and energy underpin emotion models; lyrics offer semantic insight; and context and demographics ground the recommendations in real-world usage.

Despite this progress, several open challenges remain. Many current systems still struggle with cold-start scenarios, balancing fairness with personalization, and adapting to user intent in real time. Moreover, interpretability is often lacking, as deep learning models may output recommendations without clear explanations. As music consumption becomes increasingly dynamic and individualized, future work must continue to bridge technical advances with psychological and cultural understanding. Addressing these challenges will be crucial to building recommendation systems that are not only accurate but also engaging, fair, and reflective of the diverse ways people experience music.

\bibliography{main}
\bibliographystyle{tmlr}

\end{document}